\documentclass[10pt,twocolumn,letterpaper]{article}

\usepackage{cvpr}
\usepackage{times}
\usepackage{epsfig}
\usepackage{graphicx}
\usepackage{amsmath}
\usepackage{amssymb}
\usepackage{xcolor}
\usepackage{multirow}
\usepackage{multicol}
\usepackage{subfigure, enumitem}
\usepackage{threeparttable, booktabs}

\usepackage[breaklinks=true,bookmarks=false]{hyperref}

\cvprfinalcopy % *** Uncomment this line for the final submission

\def\cvprPaperID{****} % *** Enter the CVPR Paper ID here

\newcommand{\tabincell}[2]{\begin{tabular}{@{}#1@{}}#2\end{tabular}}  
\begin{document}

%%%%%%%%% TITLE
\title{OpenDVC:\\ An Open Source Implementation of the DVC Video Compression Method}

\author{Ren Yang\\Computer Vision Laboratory\\ETH Z\"urich, Switzerland 
\\{\tt \small ren.yang@vision.ee.ethz.ch}
\and
Luc Van Gool\\KU Leuven, Belgium\\ETH Z\"urich, Switzerland 
\\{\tt \small vangool@vision.ee.ethz.ch}
\and
Radu Timofte\\Computer Vision Laboratory\\ETH Z\"urich, Switzerland 
\\\hspace{-.1em}{\tt\small radu.timofte@vision.ee.ethz.ch}
% \and
% {\tt\small \{ren.yang, vangool, radu.timofte\}@vision.ee.ethz.ch}
}

\maketitle

\section{Introduction}

We introduce an open source Tensorflow~\cite{abadi2016tensorflow} implementation of the Deep Video Compression (DVC)~\cite{lu2019dvc} method in this technical report. DVC~\cite{lu2019dvc} is the first end-to-end optimized learned video compression method, achieving better MS-SSIM performance than the Low-Delay P (LDP) very fast setting of x265 and comparable PSNR performance with x265 (LDP very fast). At the time of writing this report, several learned video compression methods~\cite{djelouah2019neural,habibian2019video,liu2019learned,yang2020heirarchical,yang2020recurrent} are superior to DVC~\cite{lu2019dvc}, but currently none of them provides open source codes. We hope that our OpenDVC codes are able to provide a useful model for further development, and facilitate future researches on learned video compression. Different from the original DVC, which is only optimized for PSNR, we release not only the PSNR-optimized re-implementation, denoted by OpenDVC (PSNR), but also the MS-SSIM-optimized model OpenDVC (MS-SSIM). Our OpenDVC (MS-SSIM) model provides a more convincing baseline for MS-SSIM optimized methods, which can only compare with the PSNR optimized DVC~\cite{lu2019dvc} in the past. The OpenDVC source codes and pre-trained models are publicly released at \url{https://github.com/RenYang-home/OpenDVC}.

\section{Implementation}

\begin{figure}[!t]
\centering
\includegraphics[width=1\linewidth]{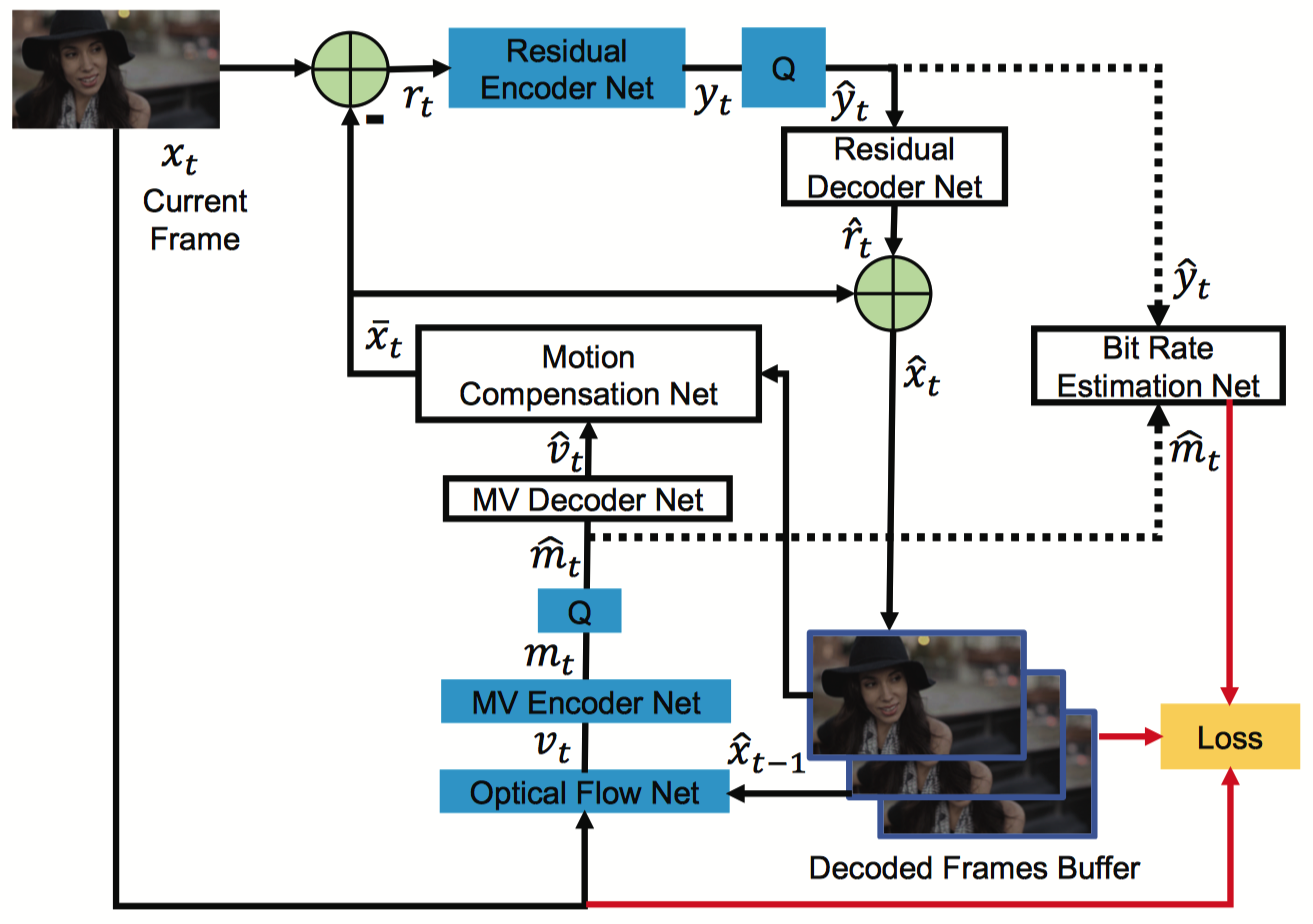}
\caption{The high-level framework of DVC~\cite{lu2019dvc}.} \label{fig:overview}
\end{figure}

In this section, we describe the implementation of our OpenDVC, which follows the framework of DVC~\cite{lu2019dvc} shown in Figure~\ref{fig:overview}. The high-level architecture of DVC is motivated by the handcrafted video coding standards~\cite{wiegand2003overview, sullivan2012overview}, \ie, adopting motion compensation to reduce the temporal redundancy and using two compression networks to compress the motion and residual information, respectively. In the following, we introduce the OpenDVC implementation of each module presented in Figure~\ref{fig:overview}.

% \subsection{Motion estimation}

\textbf{Motion estimation.} DVC utilizes the pyramid network~\cite{ranjan2017optical} to estimate the motion between the current frame and the previous compressed frame, shown as the ``Optical Flow Net'' module in Figure~\ref{fig:overview}. The large receptive field of pyramid architecture benefits DVC to handle large motions. In OpenDVC, the motion estimation network is implemented by Tensorflow in the file \texttt{motion.py}, based on a PyTorch implementation \cite{pytorch-spynet} of the pyramid network~\cite{ranjan2017optical}. We follow the settings described in \cite{ranjan2017optical} to use a 5-level pyramid network. Each level has five convolutional layers with the kernal size of $7\times7$, and with the filter numbers of 32, 64, 32, 16 and 2, respectively. As Figure~\ref{fig:overview} shows, the estimated motion $v_t$ is output from the pyramid network, denoted as \texttt{flow\_tensor} in \texttt{OpenDVC\_test\_video.py}.

\begin{figure*}[!t]
\centering
\includegraphics[width=1\linewidth]{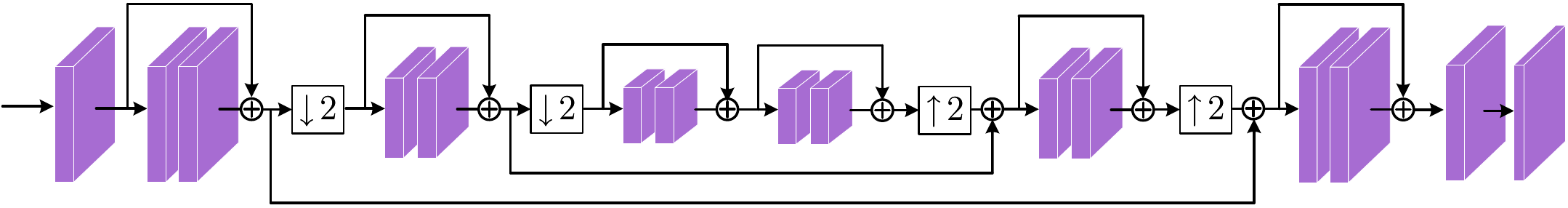}
\vspace{.5em}
\caption{The architecture of the motion compensation network~\cite{lu2019dvc,yang2020heirarchical}.} \label{fig:mc}

\end{figure*}

\textbf{Motion compression.} We follow \cite{lu2019dvc} to use the auto-encoder of \cite{balle2017end} to compress the the estimated motion. The encoder part consists of four convolutional layers with $\times2$ down-sampling, and the first three layers use the activation function of GDN~\cite{balle2017end}. In the decoder part, there are four corresponding convolutional layers with $\times2$ up-sampling, and the first three layers use the activation function of the inverse GDN~\cite{balle2017end}. In motion compression, we set the filter size as $3\times3$ and the filter number as 128 for all layers except the last layer in decoder, which has the filter number of 2 to reconstruct the 2-channel motion vector. The encoder and decoder for motion compression are implemented as the functions \texttt{MV\_analysis} and \texttt{MV\_synthesis} in \texttt{CNN\_img.py}. Different from DVC~\cite{lu2019dvc} which employs the \textit{hyperpripr} entropy model~\cite{balle2018variational}, our OpenDVC uses the \textit{factorized} entropy model~\cite{balle2017end}\footnote{\url{https://github.com/tensorflow/compression/releases/tag/v1.0}}. As such, OpenDVC has lower requirements on the input resolution, \ie, DVC requires the input height and width to be the multiples of 32, while our OpenDVC only needs height and width to be the multiples of 16. More importantly, replacing the hyperprior model with the factorized model does not lead to obvious drop of performance (please refer to Section~\ref{exp}).

\textbf{Motion compensation.} As described in DVC~\cite{lu2019dvc}, the reference frame is first warped by the compressed motion \texttt{flow\_hat}, and the motion compensation network takes as inputs the reference frame \texttt{Y0\_com}, the warped\footnote{In OpenDVC, we use the backward warping, which is implemented as \texttt{tf.contrib.image.dense\_image\_warp} in Tensorflow 1.12.} reference frame \texttt{Y1\_warp} and the compressed motion \texttt{flow\_hat} to generate the motion compensated frame \texttt{Y1\_MC}. The motion compensation network in our OpenDVC follows the architecture shown in the Appendix\footnote{\url{https://arxiv.org/abs/1812.00101}.} of \cite{lu2019dvc}. The detailed network is show in Figure~\ref{fig:mc}, in which all layers have the filter size of $3\times3$. The filter number of each layer is set to 64, except the last layer whose filter number is 3. $\uparrow2$ and $\downarrow2$ indicate up- and down-sampling with the stride of 2, respectively, and $\bigoplus$ denotes the element-wise addition.

% the skip layers are with the filter size of $1\times1$ to make the channel numbers the same for the element-wise addition operation $\bigoplus$. Other layers have the filter size of $3\times3$, and the filter numbers are shown in Figure~\ref{fig:mc}.

\textbf{Residual compression.} After motion compensation, the residual can be obtained as the difference between the compensated reference frame and the current raw frame. In our OpenDVC, we compress residual with the same method as the motion compression. The only difference is that we use the filters with the size of $5\times5$ in the auto-encoder for residual compression, instead of $3\times3$ in motion compression. The reason is that residual contains more information and consumes more bit-rate than motion \cite{lu2019dvc}, and larger filter size improves the representation ability of the auto-encoder. Finally, the reconstructed compressed frame can be obtained by adding the residual to the compensated reference frame. 

\section{Training}

In this technical report, we use the same notations as DVC~\cite{lu2019dvc}, shown in Figure~\ref{fig:overview}. The definition of the notations and their corresponding variable names in our OpenDVC codes \texttt{OpenDVC\_test\_video.py} are listed in Table~\ref{tab:notation}. 

\begin{table}[!b]
\centering
\small
\caption{Corresponding notations and variable names.}\label{tab:notation}
    \begin{tabular}{ccc}
    \cmidrule[\heavyrulewidth]{1-3}
       Definition  & Notation & Variable name\\
    \cmidrule[\heavyrulewidth]{1-3}
       Reference frame  & $\hat x_{t-1}$ & \texttt{Y0\_com}\\
    \cmidrule{1-3}
       Current frame  & $x_{t}$ & \texttt{Y1\_raw}\\
    \cmidrule{1-3}
       Estimated motion  & $v_{t}$ & \texttt{flow\_tensor}\\
    \cmidrule{1-3}
    \tabincell{c}{Latent representation\\ of motion}  & $m_{t}$ & \texttt{flow\_latent}\\
    \cmidrule{1-3}
    \tabincell{c}{Quantized latent \\representation of motion}  & $\hat m_{t}$ & \texttt{flow\_latent\_hat}\\
    \cmidrule{1-3}
    Compressed motion  & $\hat v_{t}$ & \texttt{flow\_hat}\\
    \cmidrule{1-3}
    % \tabincell{c}{Reference frame warped \\by compressed motion}  & $x^w_{t}$ & \texttt{Y1\_warp}\\
    % \cmidrule{1-3}
    \tabincell{c}{Motion compensated \\ reference frame}  & $\overline x_{t}$ & \texttt{Y1\_MC}\\
    \cmidrule{1-3}
    Residual  & $r_{t}$ & \texttt{Res}\\
    \cmidrule{1-3}
    \tabincell{c}{Latent representation\\ of residual}  & $y_{t}$ & \texttt{res\_latent}\\
    \cmidrule{1-3}
    \tabincell{c}{Quantized latent \\representation of residual}  & $\hat y_{t}$ & \texttt{res\_latent\_hat}\\
    \cmidrule{1-3}
    Compressed residual  & $\hat r_{t}$ & \texttt{Res\_hat}\\
    \cmidrule{1-3}
    Compressed frame  & $\hat x_{t}$ & \texttt{Y1\_com}\\
    \cmidrule[\heavyrulewidth]{1-3}
    \end{tabular}
\end{table}

\begin{figure*}[!t]
\centering
\subfigure{\includegraphics[width=.35\linewidth]{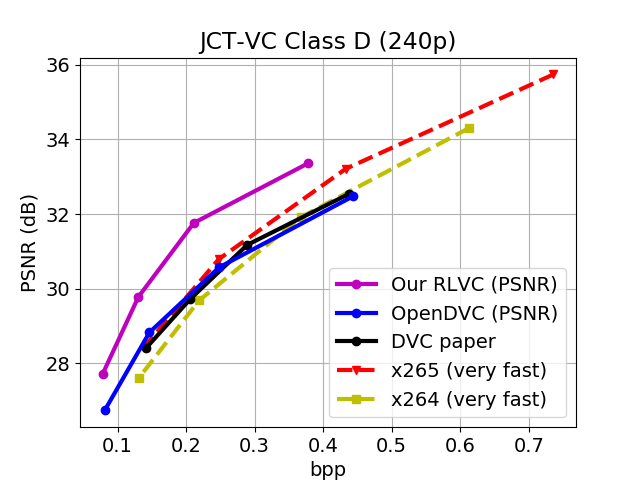}} 
\hspace{-1.8em}
\subfigure{\includegraphics[width=.35\linewidth]{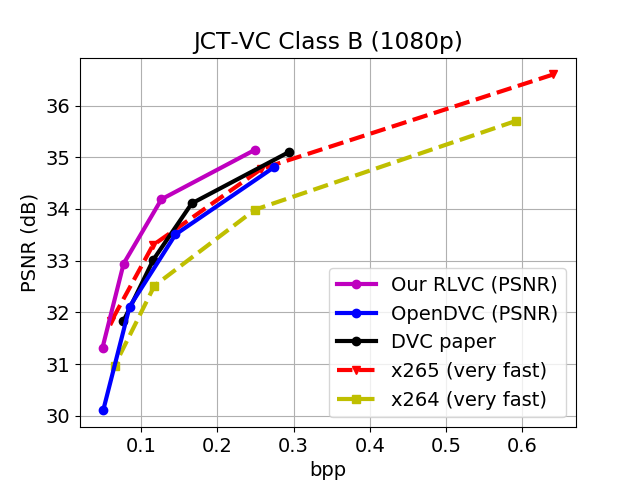}}
\hspace{-1.8em}
\subfigure{\includegraphics[width=.35\linewidth]{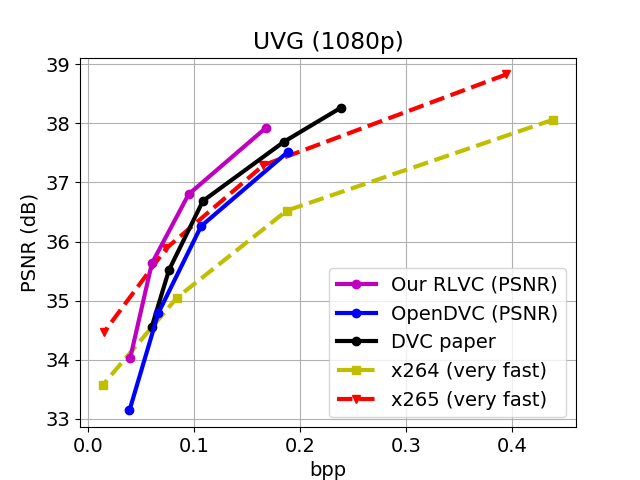}}\\
\subfigure{\includegraphics[width=.35\linewidth]{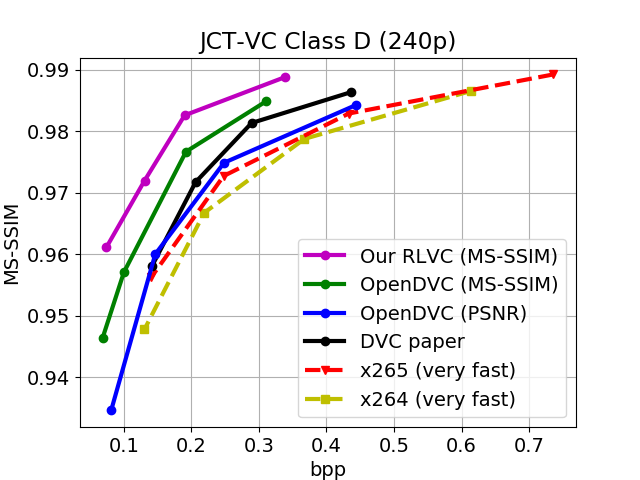}} 
\hspace{-1.8em}
\subfigure{\includegraphics[width=.35\linewidth]{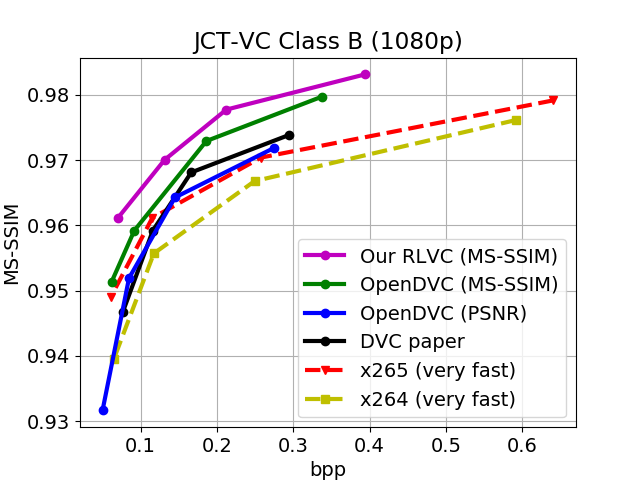}}
\hspace{-1.8em}
\subfigure{\includegraphics[width=.35\linewidth]{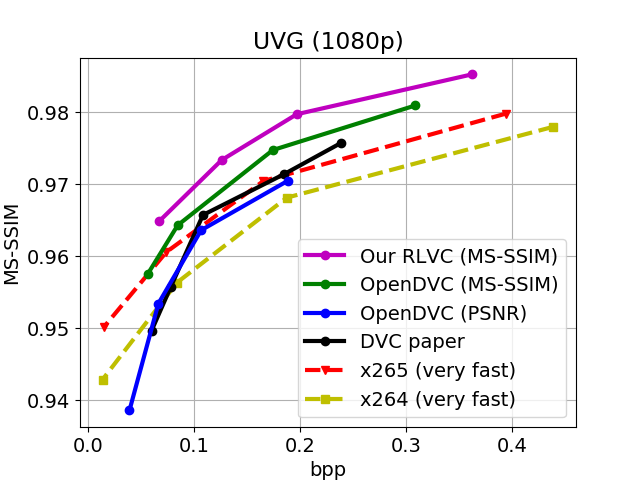}}
\vspace{.001em}
\caption{The performance of DVC~\cite{lu2019dvc}, OpenDVC and our latest RLVC approach~\cite{yang2020recurrent}.} \label{fig:curve}
\end{figure*}

The OpenDVC network is trained on the Vimeo-90k~\cite{xue2019video} dataset in a progressive manner. At the beginning, the motion estimation network is first trained with the loss function of
\begin{equation}\label{loss_me}
    \mathcal{L}_{\text{ME}} = D(x_{t},W(\hat x_{t-1}, v_t)),
\end{equation}
where $W$ is the backward warping operation. 
After the convergence of the motion estimation network, we further include the motion compression network into training with the loss including the distortion of the reference frame warped by the compressed motion and the bit-rate for compressing $\hat m_{t}$, \ie,
\begin{equation}\label{loss_mcom}
    \mathcal{L}_{\text{M}} = \lambda\cdot D(x_{t},W(\hat x_{t-1}, \hat v_t)) + R(\hat m_{t}),
\end{equation}
in which $\lambda$ balances the penalties of rate and distortion, and $R$ stands for the bit-rate estimated by the entropy model~\cite{balle2017end}. Then, the motion compensation network is trained by
\begin{equation}\label{loss_mc}
    \mathcal{L}_{\text{MC}} = \lambda\cdot D(x_{t},\overline x_{t}) + R(\hat m_{t}).
\end{equation}
When $\mathcal{L}_{\text{MC}}$ is converged, the whole network is jointly trained in an end-to-end manner, using the loss of
\begin{equation}\label{loss_all}
    \mathcal{L} = \lambda\cdot D(x_{t},\hat x_{t}) + R(\hat m_{t}) + R(\hat y_{t}).
\end{equation}
The learning rate is initially set as $10^{-4}$ for all loss functions \eqref{loss_me}, \eqref{loss_mcom}, \eqref{loss_mc} and \eqref{loss_all}. When training the whole network by the final loss of \eqref{loss_all}, the learning rate decreases by the factor of 10 after convergence until $10^{-6}$.

In OpenDVC, we first follow DVC~\cite{lu2019dvc} to train the PSNR-optimized model with the distortion $D$ as the Mean Square Error (MSE) and $\lambda=$ 256, 512, 1024 and 2048. Then, the MS-SSIM models are fine-tuned only using the final loss function \eqref{loss_all} with $D = 1 - \text{MS-SSIM}$. The MS-SSIM models with $\lambda=$ 8, 16, 32 and 64 are fine-tuned from the pre-trained PSNR models with $\lambda=$ 256, 512, 1024 and 2048, respectively. Note that, we use BPG~\cite{BPG} to compress the I-frames for the PSNR models in OpenDVC, and use the learned image compression method~\cite{lee2019context} to compress the I-frames for the MS-SSIM models. Specifically, the BPG with QP = 37, 32, 27 and 22 is used for PSNR models with $\lambda=$ 256, 512, 1024 and 2048, respectively. The MS-SSIM models with $\lambda=$ 8, 16, 32 and 64 use \cite{lee2019context} with the quality levels of 2, 3, 5 and 7, respectively.

\section{Performance} \label{exp}

The rate-distortion performance of OpenDVC is demonstrated in Figure~\ref{fig:curve}, in comparison with the results reported in DVC~\cite{lu2019dvc}. It can be seen that the OpenDVC (PSNR) model achieves comparable performance with DVC in terms of PSNR and MS-SSIM, and OpenDVC (MS-SSIM) obviously outperforms DVC in terms of MS-SSIM. Note that, Figure~\ref{fig:curve} directly uses the results of DVC, x265 (very fast) and x264 (very fast) reported in \cite{lu2019dvc}.

\section{Our latest works}

In 2020, we proposed a Hierarchical Learned Video Compression (HLVC) approach~\cite{yang2020heirarchical} with hierarchical quality and recurrent enhancement layer. Our HLVC approach is published in CVPR 2020. The paper can be downloaded at  \url{https://arxiv.org/abs/2003.01966}, and the project page is at \url{https://github.com/RenYang-home/HLVC}.

Later, we proposed a Recurrent Learned Video Compression (RLVC) approach~\cite{yang2020recurrent} with recurrent auto-encoder and recurrent probability model. The paper is publicly available at \url{https://arxiv.org/abs/2006.13560}. The results of our latest RLVC~\cite{yang2020recurrent} approach are also illustrated in Figure~\ref{fig:curve}, which clearly outperforms the performance of DVC and also advances the state-of-the-art of learned video compression approaches (refer to our paper).

{\small
\bibliographystyle{ieee_fullname}
\bibliography{egbib}
}

\end{document}